\documentclass{article}

\usepackage{arxiv_2}

\usepackage[utf8]{inputenc} 
\usepackage[T1]{fontenc}    
\usepackage{hyperref}       
\usepackage{url}            
\usepackage{booktabs}       
\usepackage{amsfonts}       
\usepackage{nicefrac}       
\usepackage{microtype}      
\usepackage{lipsum}		
\usepackage{graphicx}
\usepackage{natbib}
\usepackage{doi}

\usepackage{multirow}
\RequirePackage{natbib}
\hypersetup{colorlinks,allcolors=blue}
\renewcommand{\shorttitle}{Yield forecasting based on short time series with high spatial resolution data}

\RequirePackage{amsthm,amsmath,amsfonts,amssymb}
\RequirePackage{float}

\begin{document}

\title{Yield forecasting based on short time series with high spatial resolution data}

\author{
\thanks{Corresponding author email: \it{spokal@huskers.unl.edu}}
Sayli Pokal\\[4pt] 
\textit{University of Nebraska-Lincoln, Nebraska, USA}
\\[2pt]
Yuzhen Zhou\\[4pt]
\textit{University of Nebraska-Lincoln, Nebraska, USA}
\\[2pt]
Trenton Franz\\[4pt]
\textit{University of Nebraska-Lincoln, Nebraska, USA}
\\[2pt] }


\maketitle


\begin{abstract}
{ Precision agriculture, also known as site-specific crop management, plays a crucial role in modern agriculture. Yield maps are an essential tool as they help identify the within-field variability that forms the basis of precision agriculture. If farmers could obtain yield maps for their specific site based on their field's soil and weather conditions, then site-specific crop management techniques would be more efficient and profitable. However, forecasting yield and producing reliable yield maps for an individual field can be challenging due to limited historical data. This paper proposes a novel two-stage approach for site-specific yield forecasting based on short-time series and high-resolution yield data. The proposed approach was successfully applied to predict yield maps at three different sites in Nebraska, demonstrating the method's ability to provide fine resolution and accurate yield maps.}

\keywords{precision agriculture; site-specifc yield forecasting; yield maps; clustering; auto-regressive; Gaussian copula}

\end{abstract}

\graphicspath{ {./Images/} }

\section{Introduction}
 
Precision agriculture, also known as site-specific crop management, plays a crucial role in modern agriculture \citep{pedersen2017precision}. It involves using spatio-temporal data to identify variability within a field and adjust crop treatments accordingly. Recent advancements in technology, such as yield monitors and Global Positioning Systems (GPS), have made it possible to collect yield data at geo-referenced points and create yield maps that visualize the variability within the field. However, yield maps can differ significantly from year to year due to factors such as weather, crop diseases, and crop management techniques. Predicting future yields is, therefore, challenging and requires analyzing years of yield maps, soil data, weather patterns, and crop management information.

Accurate prediction of crop yield is pivotal in aiding farmers and policymakers to make well-informed decisions regarding soil, crop management, marketing, and storage. Two common approaches for estimating crop yield include implementing process-based crop simulation models and statistical analysis of spatio-temporal data sets. Crop simulation models are mechanistic models that consider the crop’s physiological characteristics and various environmental factors \citep{wang2002development, holzworth2014apsim, basso2013review}. However, these models can be challenging to implement as they require a large number of parameters and input variables that are not always readily available. On the other hand, statistical models rely on past observational data to identify relationships and patterns in the data that can be applied to predict future crop yields. Although statistical models do not directly incorporate plant growth mechanisms, they can be useful in forecasting crop yield.

There are several statistical models available for forecasting yield at a large scale (regional or state level), as discussed in previous studies (\citep{newlands2014integrated, bussay2015improving, paudel2021machine}). However, only a few approaches are available for forecasting yield for an individual farm or site-specific yield forecasting. For any given year, \cite{site1} predicts the yield within a site using methods such as stepwise multiple linear regression, projection pursuit regression, and neural networks. Other techniques, including Bayesian networks, regression trees, and artificial neural networks (ANN), have also been used for site-specific yield prediction. However, predicting site-specific yield for future years is more challenging than predicting yield within an individual site year. 

The most common approaches for site-specific yield forecasting include fitting linear regression models or spatial econometric models, as demonstrated by \citep{site2, anselin2004spatial, schwalbert2018forecasting}. \cite{anselin2004spatial} used yield monitor data to obtain site-specific yield forecasts by implementing a spatial econometric model. \cite{site2} and \cite{ schwalbert2018forecasting} used high-resolution satellite imagery data at the mid-growing season to identify within-field variability and used ordinary least square regression and spatial econometric models to forecast site-specific yield at the end of the season.  \cite{lambert2004comparison} compared four spatial regression models that incorporate spatial correlation in the economic analysis of variable rate technology. \cite{li2016estimating} estimated site-specific crop yield response functions using varying coefficient models and developed a decision system that provides input prescriptions for producers. The existing methods for forecasting site-specific yield are either based on satellite imagery data or data with many time points. However, not much literature exists for forecasting yield based on historical yield maps for a short time series and high dimensional spatial data.

In this paper, we aim to address this gap in the field and develop a new approach for site-specific yield forecasting based on historical yield maps. Our study was motivated by a maize yield data set from a field in Mead, Nebraska, consisting of 7 years of historical yield maps. Each yield map has a spatial resolution of $10m$, and the field size is $800m \times 800m$. Our goal is to estimate yield for the future growing season and predict the spatial pattern of yield distribution in the field, i.e., obtain yield maps. However, forecasting yield with the data at hand is challenging due to the short time series and noise in the high-resolution data. Existing methods for site-specific yield forecasting are not designed to handle such short-time series data with only 7 time points; hence, a new approach is required.

The spatially varying auto-regressive (SVAR) model employed by \cite{Shand2018} is a model with the potential to handle short-time series spatial data. \cite{Shand2018} used the model to predict HIV diagnosis rates in US states based on county-level HIV data, where the data was abundant in space but included only a few time points. Instead of creating a time series model for each county, the SVAR model jointly modeled the parameters for each county using a Gaussian copula. The SVAR model uses a spatial dependence structure that combines information from its neighborhood time series. The neighbors essentially act as “replicates”, making the model’s forecasts more reliable compared to the single time series approach.

However, the SVAR model was developed for county-level or lattice spatial data and cannot be directly applied to the yield data, which is continuous in nature. Due to the high dimensionality of the yield data, directly implementing the SVAR model on the yield data would be computationally expensive, with too many parameters to estimate. Additionally, the high-resolution data tends to be noisy and may result in lower prediction accuracy if fed directly into the model. To make the model forecasts reliable and the number of parameters reasonable, there is a need to reduce the dimension of the data and the noise. A common approach to address this issue in spatial statistics is to divide the field into blocks and aggregate the data within the block. However, this approach works well only for homogeneous spatial fields. For an inhomogeneous field, the blocking approach would eliminate the fine patterns in the data, leading to inaccurate yield predictions.

In this paper, we propose a novel two-stage approach for site-specific yield forecasting based on short-time series and high-resolution spatial data that addresses the above issues. In the first stage, we develop a clustering approach for dimension reduction and noise reduction, which retains the fine pattern of the spatial field. In the second stage, we apply a modified version of the SVAR model to obtain yield forecasts for the future growing season. Implementing the proposed method at three different sites in Nebraska, we demonstrate that our method provides finer resolution and more accurate yield maps than the existing approach. The proposed method can thus help implement more effective site-specific management strategies.

The rest of the paper is organized as follows: Section \ref{sec: data and site description} describes the data and site. The details of the proposed methods are presented in Section \ref{sec: Methods}. In Section \ref{sec: Results}, we implement the proposed model with maize yield data and compare it with existing models. Section \ref{sec: conclusions} concludes the study. Technical details and results for two other independent sites are included in the Appendix.

\section{Data and Site Description}
\label{sec: data and site description}

The data were collected at three different sites in Nebraska, namely Mead, Brule, and Site 6, located on the University of Nebraska research and extension farms. The size of each field was approximately $64 ha$ ($800 m \times 800 m$). Historical yield maps for at least seven years were available for each site, along with corresponding hydro-geophysical maps. 
Yield data was available for every $100 m^2$ of the field, with historical yield maps having a spatial resolution of $10 m$. Spatial maps of shallow and deep electrical conductivity (EC) and soil water content (SWC) with a $10 m$ resolution were also available for each field, based on the hydro-geophysical surveys conducted in the field. EC and SWC geophysical maps were pre-processed using empirical orthogonal function (EOF) analysis. The EOF analysis decomposes the observed SWC and EC variability measured by the hydro-geophysical surveys into a set of orthogonal spatial patterns (EOFs), which are invariant in time, and a set of time series expansion coefficients (ECs) which are invariant in space \citep{perry2007}. Using EOFs helps reduce individual survey noise and instrument error while preserving the dominant geophysical spatial patterns. For yield forecasting analysis, the first EOFs of the SWC and EC geophysical maps were used as predictors, along with relative elevations of the site. Complete details of the multivariate statistical EOF analysis can be found in \citep{perry2007} and \cite{korres2010}. See \cite{franz2020role} for a complete description of the crop yield and geophysical datasets. 

Crop rotation was done at the Mead site, where maize and soybean are grown in alternate years. Maize was planted at Brule and Site 6 during all years. Planting occurred between late April and early May, depending on the field and weather conditions. Irrigation was provided as required, starting mid-June through early September. Irrigation, herbicide, pesticide, and other crop management practices followed the standard best management practices prescribed for production-scale maize systems. However, information on these technicalities was not available for all sites and years.

Weather information for each site was collected from the nearest Nebraska Mesonet Station within a $20 km$ radius. Data was collected on rainfall totals (RT) and potential evaporation (PET) for each growing season, May-June (vegetation growth), July-August (reproduction/grain filling), and seasonal totals (May-September). Rainfall total is known to be a good predictor of average annual yield. PET provides information about the crop water demand, accounting for factors such as temperature, wind speed, and solar radiation. Information on the average storm depth (SD) and average inter-storm arrival rates (SA) is also included for each year. Table \ref{table: site description} provides details of each site.

\begin{table}[H]
\caption{Study Site Description} 
\centering 
\scalebox{0.88}{
\renewcommand{\arraystretch}{2}
\begin{tabular}{p{1.2cm} p{3cm} p{1.9cm} c } 
\hline\hline  
\textbf{Study Site} & \textbf{Landuse} & \textbf{Mesonet Station} & \textbf{Crop yield years}   \\ [0.5ex] 
\hline 
Mead   & Rainfed maize and soybean rotation & Ithaca 3E & 2001 - 2017 \\
Brule  & Irrigated maize & Big Springs 8NE & 2010 - 2016  \\
Site 6 & Irrigated maize & Ithaca 3E & 2001 - 2017     \\     \\
\hline
\end{tabular}
}
\label{table: site description}
\end{table}

\section{Methods}
\label{sec: Methods}
The goal of this paper is to obtain one-year ahead forecasts for site-specific maize yield and to obtain yield maps of the field. We propose a model for forecasting yield at an individual farm level using information such as geophysical variables (soil water content and EC), weather conditions, relative elevation, and historical yield data of the field. The challenge with obtaining forecasts for the data at hand is that we have short time series data that is noisy and high dimensional. Implementing forecasting models directly on the data is not feasible and does not provide good forecasts. We thus propose a novel two-stage approach to obtain forecasts for a short time series and high-resolution spatial data. First, we develop a clustering approach for data aggregation. Second, we implement a modified version of the SVAR forecasting model for obtaining yield forecasts for the future growing season.

\subsection{Data aggregation}

\subsubsection{Blocking approach}
 
The idea behind spatial data aggregation is that two points that are close to each other in space are correlated. Aggregating these data points will help reduce noise and dimension while retaining helpful information. One way to perform spatial data aggregation is to divide the entire field into equal-sized blocks and average all the observations within each block. Averaging the observations within a block results in a single value corresponding to each block in the field. This method is referred to as data aggregation by blocking.

\begin{figure}[H]
\centering
  \includegraphics[scale=0.35]{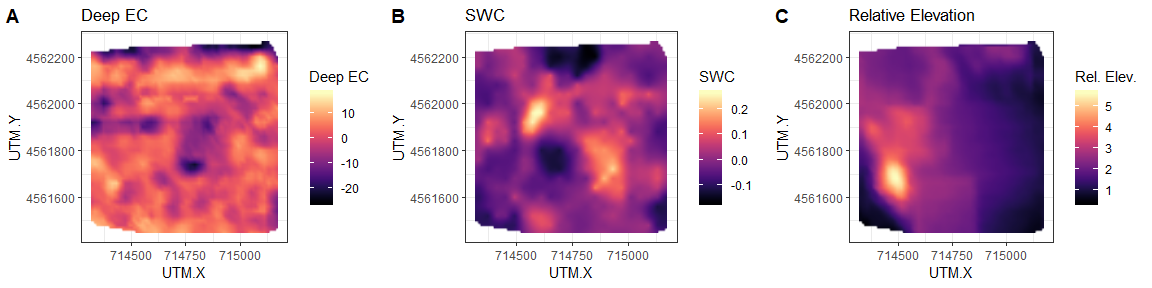}
  \caption{Maps of the spatial covariates at site in Mead, A: Map of Deep EC, B: Map of Soil Water Content, C: Map of Relative Elevation. Universal Transverse Mercator (UTM) is a map projection system with units of meters.}
  \label{fig:predictor_maps}
\end{figure}

Aggregating data within a region is helpful if the observations are homogeneous. However, the observations within a given region may not be homogeneous with respect to all the covariates. Looking at Figure \ref{fig:predictor_maps}, we observe that the geophysical maps for each covariate show different spatial patterns. If we define homogeneous regions within the field based on relative elevation, the corresponding regions for Deep EC may not contain homogeneous observations. In this situation, blocking will lead to information loss for Deep EC. 

Suppose we divide the field into 64 equal-sized blocks and aggregate the data within each block. The observations within a block will not necessarily be homogeneous for all variables. Aggregating the data will not be efficient as it will eliminate the fine patterns in the field. Thus, data aggregation by blocking may lead to the loss of useful information. If we divide the field into smaller blocks, we might not achieve significant noise reduction, and we will also require the estimation of a larger number of model parameters. Therefore, there is a trade-off between the number of parameters to be estimated and avoiding over-aggregation.

\subsubsection{Clustering approach}

In the case of an inhomogeneous field, an approach for reducing noise would be to identify similar observations and aggregate these similar observations, leading us to the clustering approach. Grouping observations that are similar and aggregating over these groups will reduce the noise in the data. It will also help identify patterns in the field, such as regions corresponding to low and high yields. The clustering approach will help retain the fine patterns in the EC map, whereas the blocking approach will not retain these fine patterns in the data. Before implementing clustering, we need to specify the variables used for clustering and the number of clusters. 

\subsubsection*{Variables used for clustering}

Ideally, we want to create groups that identify patterns in the response variable, i.e., the yield, as it will help identify within-field variability in the yield data. To achieve this, we need to identify variables correlated to the response and use these variables to perform clustering. It has been shown that there is a correlation between soil EC and yield, where lower values of soil EC are correlated to lower yield values \citep{ecyield}. We also observed from the historical yield data that the future growing season yield is correlated to the current yield. Hence, historical yield and EC should be good predictors of within-field variability. The distribution of soil EC may stay the same over the years. However, the distribution of yield may be different from year to year. In practice, we suggest using variables correlated to the response or historically known to be good predictors of the response as clustering variables. 

\subsubsection*{Number of clusters}

Since the purpose behind clustering is to find homogeneous groups of observations in the data that help predict the response, these clusters need not represent the actual clusters in the data. The number of clusters will depend on what seems reasonable for the data set. Using a large number of clusters may not help with noise reduction and will require estimating a large number of parameters. In contrast, a small number of clusters may result in over aggregation of data and loss of useful information. We need to select the number of clusters that will lead to a reasonable number of parameters to be estimated while retaining valuable information for forecasting yield. One can consider cluster validity statistics, such as a scree plot, as a starting point for selecting the number of clusters. We suggest trying a range of values from small to large and selecting the value that provides good forecasts. 

\subsubsection*{Implementation} 

We implemented k-means clustering using Deep EC and the current year's yield as clustering variables. Initially, we tried several models using different combinations of covariates and historical yield to determine the clustering variables that result in the best forecasting performance. We found that the model with Deep EC and the current year's yield gave the best forecasting results, which supported our intuition behind using these variables to perform clustering. We performed clustering for cluster sizes of 25 and 64, which allowed us to compare the forecasting performance of the models using clustering versus blocking as a data aggregation method.
Once we obtained the blocks/clusters for the field, data aggregation was performed by taking the arithmetic average of all the observations within each block/cluster. To reduce the skewness in the yield data, we log-transformed the yield and then aggregated the data within each cluster. As a result, we end up with one value per block/cluster for the log-transformed yield and the covariates.

\subsection{The forecasting model}

\subsubsection{Trend analysis}
 
The maize yields show an increasing trend over time. The increase in yield over the years is primarily due to the advancement in technology and the use of hybrid seeds. The trend can be significantly affected by climatic conditions such as drought or individual storms (e.g., damage from wind or hail). Thus, weather information needs to be incorporated while fitting the trend model. We fitted a linear trend model for our data, considering the average log-transformed yield within the field as the response. We included predictors such as year, rainfall total (RT), potential ET (PET), storm depth rate (SD), and storm arrival rate (SA) for the season of July through August. Model selection was performed to select the best model for each site corresponding to the lowest mean squared error. The trend model selected for each site is summarized in the table below.

\begin{table}[H]
\caption{Trend model} 
\centering 
\scalebox{0.9}{
\renewcommand{\arraystretch}{2}
\begin{tabular}{p{1.7cm} p{8cm} } 
\hline\hline  
\textbf{Site} & \textbf{Trend model} \\ [0.5ex] 
\hline 
Mead &   $log(yield) = year + RT + PET + SD$ \\
Brule &  $log(yield) = year + RT + PET$      \\
Site 6 & $log(yield) = year + RT + PET$      \\ 
\hline
\end{tabular}
}
\end{table}

The fitted trend model was then used to de-trend the yield at each site, and normalized yield data was obtained by subtracting the average log-transformed yield from the aggregated data. Validating the trend model is challenging in the case of short time series. We suggest adopting a domain expert's knowledge of the topic or using trend models established in the past.

\subsubsection{Bayesian forecasting model}

We implemented a modified version of the SVAR model to obtain forecasts for the short-time series. Let $Y_{i,t}$ and $Z_{i,t}$ denote the aggregated yield and the normalized yield, respectively, for the $i$th cluster ($i=1, 2, \dots, n$) at time $t$ ($t= 1, 2,\dots, T$). We model $Z_{i, t}$ using Bayesian hierarchical model as follows, 
\begin{equation}
Z_{i,t}= \rho_iZ_{i,t-1}+\epsilon_{i,t},
\end{equation}
where $\epsilon_{i,t} \stackrel{i.i.d.}{\sim} N(0, \sigma^2)$, and $\rho_i \in (-1, 1)$, the  AR(1) coefficient of the $i$th cluster. The prior of $\boldsymbol{\rho} = (\rho_1, ..., \rho_n)^\top$ is modeled using a Gaussian copula with covariance matrix $\boldsymbol{\Omega}$. The copula approach allows modeling the dependence structure among $\rho_i$s while providing the flexibility of choosing appropriate marginal distributions for each $\rho_i$. Hence, the estimation and inference of any single auto-correlation coefficient $\rho_i$ become more accurate and stable by combining the information from its neighborhood time series.

In \cite{Shand2018}, the covariance matrix $\boldsymbol{\Omega}$ is modeled using CAR model given by \cite{leroux} with variance $ \tau^2_{\rho} $ and spatial correlation parameter $ \lambda_{\rho} $, 
\begin{equation}
 \boldsymbol{\Omega} = \tau^2_{\rho} {(1-\lambda_{\rho} \mathbf{I} + \lambda_{\rho} \mathbf{R})},
\end{equation}
where $ \mathbf{R} $ denotes the neighborhood matrix, the $i$th diagonal element of $\mathbf{R}$ represents the total number of neighbors for the group $i$, and the $(i, j)$th  off-diagonal element is $-1$ if $i$ and $j$ are neighbors and $0$ otherwise. 

When blocking is used for data aggregation, the neighborhood structure of the data is preserved. The blocks that share edges or nodes are considered neighbors. Typically, a block in the center of the field will have eight neighbors, and those on the field's border will have three to five neighbors. The SVAR model can be implemented using this neighborhood structure.

However, the neighborhood structure is not preserved when clustering is used for data aggregation. To model the dependence among $\rho_i$s, we need to redefine the ``neighbors" of a given cluster based on the cluster similarities rather than by spatial locations.   

We used the average cluster separation matrix to define cluster neighbors. Average cluster separation is defined as the matrix of mean dissimilarities between points of every pair of clusters; it is the same as the dissimilarity matrix obtained using average linkage in hierarchical clustering \cite{Nielsen2016}. For $i = 1, ..., n$, let $C_i = \{x_{i1}, x_{i2}, ..., x_{im_i}\}$ be the $i$th cluster, where $x_{ij}$ is the $j$th object in the cluster and $m_i$ is the size of the cluster. The $(i, j)$th entry of the average cluster separation matrix is defined by the average distance of all pairs of objects in these two clusters, \[D(C_i, C_j) = \frac{1}{m_i  m_j}  \sum_{k=1}^{m_i} \sum_{\ell=1}^{m_j} d(x_{ik}, x_{j\ell}),\]
where $d(\cdot, \cdot)$ is the Euclidean distance.

Based on the concept of the $\epsilon$-neighborhood graph \citep{enn}, we considered all clusters whose pairwise distances are smaller than $\epsilon$ as neighbors. Specifically, the clusters $C_i$ and $C_j$ are neighbors only if $D(C_i, C_j) < \epsilon$. Thus, the neighborhood matrix $\mathbf R$ under the clustering approach is well defined.  

The model parameters were estimated using the MCMC algorithm. The prediction for normalized yield $Z_{i,t}$ was obtained by sampling from the posterior predictive distribution using forward sampling. 
For each iteration, we have 
\begin{equation}
\hat{Z}_{i,t}= \hat{\rho}_i Z_{i,t-1}+\hat{\epsilon}_{i,t}.
\end{equation}
For more details regarding the implementation of the SVAR model, refer to \cite{Shand2018}.

Forecasts were obtained for each cluster using the above model. These forecasts were on the normalized yield data scale. The normalized yield forecasts were then back-transformed on the yield data scale by adding the predicted trend value and taking the exponential. The yield values for each spatial point location in the field were then obtained by assigning the yield value for the cluster to all the observations within the cluster. The above process allowed us to obtain fine-resoultion yield maps. 



\subsubsection*{Handling missing data}

Missing data is a common issue with many real-life data sets and can be an issue for parametric models. In the case of our data, the year 2009 is missing for all the sites. The correlation structure in the SVAR model assumes that the time difference is equally spaced; this assumption is violated since the yield for the year 2009 is missing. We provided a simple fix to handle the missing data issue. We considered the normalized yield to be approximately normally distributed, and then, using the property of conditional multivariate normal distribution, we obtained the conditional distribution of yield for the year 2009, given the observed yield data. For mathematical details, refer to Appendix A. 
The mean and variance parameters were estimated using the MCMC algorithm. A random sample was drawn from the conditional distribution of $2009$ yield given the observed yield and was considered as the yield for $2009$. The model parameters were updated with each MCMC iteration, and a random sample for the missing year was drawn from the updated conditional distribution during each iteration as well.

\section{Results}
\label{sec: Results}

We compared our proposed clustering-based SVAR model to four other models. Model $1$ is the clustering-based SVAR model; in this model we used clustering for data aggregation and SVAR model for forecasting. In model $2$ we used clustering for data aggregation and the random forest algorithm for forecasting. In model $3$ and $4$ we used blocking for data aggregation and implemented SVAR and the random forest for forecasting, respectively. Models $1$ -- $4$ were fitted using the normalized yield data as the response variable. Model $5$ did not use data aggregation but was fitted using the log-transformed, de-trended yield as the response variable; the random forest algorithm was applied for forecasting. 

The forecasting performance of these models was evaluated using prediction R-squared ($R^2$), mean squared prediction error (MSPE) and mean absolute prediction error (MAPE). Let $y_i$ and $\hat{y}_i,\ i=1, 2, \dots, n$ denote the observed yield and the predicted yield respectively corresponding to the i\textsuperscript{th} cluster/block for the year 2017. The performance metrics were defined as follows, 
\begin{align}
&R^2= 1- \frac{\sum_{i=1}^n (y_i-\hat{y}_i)^2}{\sum_{i=1}^n (y_i-\bar{y})^2}\\
&\text{MSPE}=\frac{1}{n}\sum_{i=1}^n  (y_i-\hat{y}_i)^2 \\
&\text{MAE}= \frac{1}{n}\sum_{i=1}^n  |y_i-\hat{y}_i|
\end{align}

Table \ref{Forecasting_performance_table_Mead} presents the results for the Mead site across different models for clusters of size $n=25$ and $n=64$. 


\begin{table}[H]
\caption{Forecasting performance at the Mead site} 
\label{Forecasting_performance_table_Mead}
\centering 
\scalebox{0.9}{
\renewcommand{\arraystretch}{2}
\begin{tabular}{p{1.8cm} p{1.6cm} c  c  c  c  p{1.7cm} } 
\hline\hline  
\textbf{Aggregation Method} & \textbf{Forecasting Model} & \textbf{\# Clusters}  & \textbf{$R^2\,(\%)$} & \textbf{MSPE} & \textbf{MAPE} & \textbf{Predicted Average} \\ [0.5ex] 
\hline 
Clustering & SVAR          & \multirow{2}{*}{25} & 81      & 0.141 & 0.322 & 11.717 \\
Clustering & Random Forest &                     & 74.153  & 0.192 & 0.381 & 11.739 \\
\hline
Blocking   & SVAR          & \multirow{2}{*}{25} & 28.637  & 0.316 & 0.461 & 11.693 \\
Blocking   & Random Forest &                     & 16.579  & 0.369 & 0.473 & 11.735 \\
\hline
Clustering & SVAR          & \multirow{2}{*}{64} & 77.238  & 0.285 & 0.397 & 11.718 \\
Clustering & Random Forest &                     & 65.131  & 0.436 & 0.443 & 11.738 \\
\hline
Blocking   & SVAR          & \multirow{2}{*}{64} & 29.972  & 0.434 & 0.546 & 11.703 \\
Blocking   & Random Forest &                     & 18.55   & 0.505 & 0.553 & 11.732 \\
\hline
None       & Random Forest &         -           & 28.556  & 0.827 & 0.677 & 11.726 \\ 
\hline 
\end{tabular}
}
\end{table}

Table \ref{Forecasting_performance_table_Mead} shows that the clustering-based SVAR model with $n=25$ clusters achieved the best forecasting performance. Compared to other models, this one had the highest prediction $R^2 = 0.81$ and the lowest values for MSPE and MAPE. The predicted yield from this model was 11.717 Mg/Ha, while the actual average yield in 2017 was 12.023 Mg/Ha. The clustering-based random forest model with 25 clusters performed as the second-best among all the models.

The study shows that models that used clustering consistently outperformed those that used blocking for data aggregation. We also observed that models that used the SVAR model for forecasting performed better than the random forest-based models. Model 5 (without data aggregation) performed similarly to the blocking-based SVAR models. These findings suggest that implementing clustering as a data aggregation method leads to better forecasting performance. Finally, the study found that models with 25 clusters outperformed those with 64 clusters, likely due to the increased variability in the data.

\begin{figure}[H]
\centering
  \includegraphics[scale=0.34]{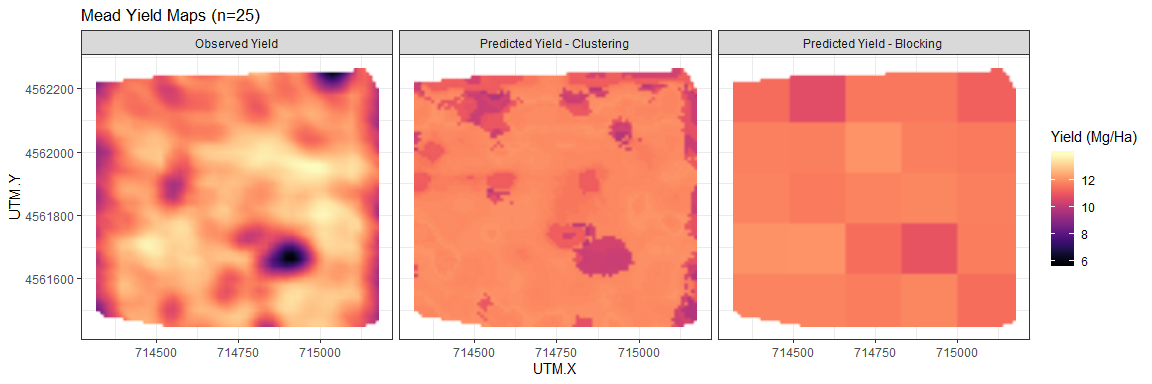}
  \caption{Yield maps at Site Mead for 25 groups, A: Observed yield map for 2017, B: Predicted yield map for 2017 using clustering based SVAR model, C: Predicted yield map for 2017 using blocking based SVAR model}
  \label{fig:Mead_yieldmaps_25}
\end{figure}

\begin{figure}[H]
\centering
  \includegraphics[scale=0.34]{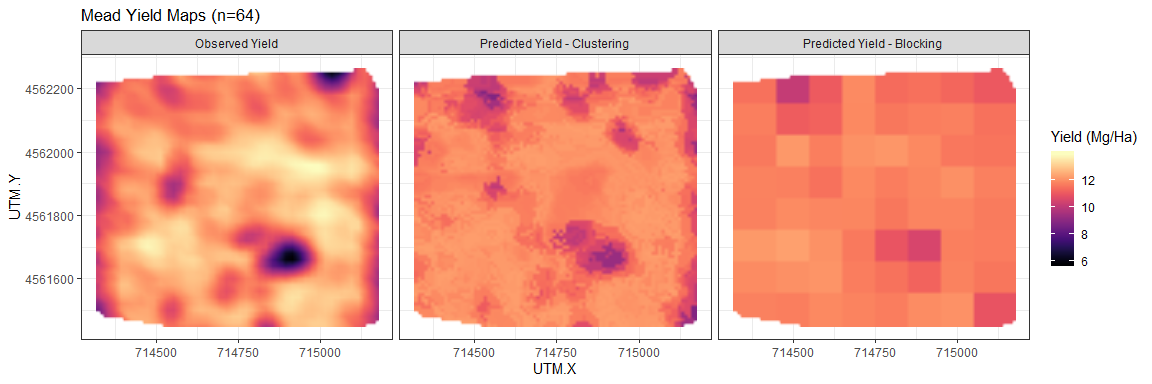}
  \caption{Yield maps at Site Mead for 64 groups, A: Observed yield map for 2017, B: Predicted yield map for 2017 using clustering based SVAR model, C: Predicted yield map for 2017 using blocking based SVAR model}
  \label{fig:Mead_yieldmaps_64}
\end{figure}

Figure \ref{fig:Mead_yieldmaps_25} and \ref{fig:Mead_yieldmaps_64} display the yield maps comparing the clustering-based SVAR and the blocking-based SVAR model for $25$ and $64$ clusters/blocks, respectively. The study found that the clustering-based SVAR model generated yield maps with finer resolutions that were more representative of the true yield maps than the blocking-based method. The clustering-based model accurately identified the pattern of yield distribution in the field and provided more information compared to the blocking-based model. The reliable yield maps produced by the clustering-based model can help understand the within-field variability, and while the clusters cannot be used as management zones by themselves, they can help develop effective management strategies.  

Furthermore, the study also examined the forecasting results for different values of $\epsilon$ in the $\epsilon$-nearest neighborhood matrix. It was found that the results were not sensitive to the values of $\epsilon$ as long as each cluster had at least one neighbor. The forecasting results for different values of $\epsilon$ are presented in Appendix B.

To verify the effectiveness of our proposed clustering-based SVAR model, we conducted the same analysis on Brule and Site 6, which are two independent sites. The results are presented in Appendix C, which includes the forecasting accuracy tables (Table \ref{Forecasting_performance_table_Brule}, \ref{Forecasting_performance_table_Site_6}) and the yield maps (Figure \ref{fig:Brule_yieldmaps_25}, \ref{fig:Brule_yieldmaps_64}, \ref{fig:Site6_yieldmaps_25}, \ref{fig:Site6_yieldmaps_64}). We found that for both sites, the clustering-based SVAR models with 64 clusters outperformed the other models and produced the most accurate yield maps. This confirms the effectiveness of our proposed model.

\section{Conclusions}
\label{sec: conclusions}

This study proposes a new approach for site-specific yield forecasting using a short time series of high-dimensional spatial data. Instead of using the common blocking approach in spatial statistics to aggregate the data, this approach uses clustering as a data aggregation technique. Clustering helps to retain the fine-scale patterns in the data and provides fine-resolution yield maps that more accurately represent the observed yield maps. The SVAR forecasting model enables the regression coefficient to vary across each cluster while also incorporating information from neighboring clusters. Furthermore, this method provides a way to obtain site-specific yield forecasts when the data has less than ten time points. The clustering-based SVAR method provides fine-resolution yield maps that more accurately represent the observed yield maps than the blocking-based models. The proposed model was validated by evaluating performance at three independent sites in Nebraska. However, due to the short time series, it was validated using a single year, i.e., 2017. Future studies can evaluate the data over multiple years across different sites. Although further research is needed to obtain management areas based on clusters, this study is one of the few studies forecasting site-specific yield for a short time series based on yield-monitor data and geophysical maps. Existing studies make use of satellite image data or use data collected over long periods of time to obtain yield forecasts. 

\section*{Acknowledgements}
We would like to thank Nathan Thorson of the Eastern Nebraska Research and Extension Center and the West Central Research and Extension Center for providing crop yield information and access to study sites.

\bibliography{ARXIV_yield}

\begin{thebibliography}{21}
\providecommand{\natexlab}[1]{#1}
\providecommand{\url}[1]{\texttt{#1}}
\expandafter\ifx\csname urlstyle\endcsname\relax
  \providecommand{\doi}[1]{doi: #1}\else
  \providecommand{\doi}{doi: \begingroup \urlstyle{rm}\Url}\fi

\bibitem[Pedersen et~al.(2017)Pedersen, Lind, et~al.]{pedersen2017precision}
S{\o}ren~Marcus Pedersen, Kim~Martin Lind, et~al.
\newblock \emph{Precision Agriculture: Technology and Economic Perspectives}.
\newblock Springer, 2017.

\bibitem[Wang et~al.(2002)Wang, Robertson, Hammer, Carberry, Holzworth, Meinke, Chapman, Hargreaves, Huth, and McLean]{wang2002development}
Enli Wang, MJ~Robertson, GL~Hammer, Peter~S Carberry, D~Holzworth, Holger Meinke, SC~Chapman, JNG Hargreaves, NI~Huth, and G~McLean.
\newblock Development of a generic crop model template in the cropping system model apsim.
\newblock \emph{European journal of Agronomy}, 18\penalty0 (1-2):\penalty0 121--140, 2002.

\bibitem[Holzworth et~al.(2014)Holzworth, Huth, deVoil, Zurcher, Herrmann, McLean, Chenu, van Oosterom, Snow, Murphy, et~al.]{holzworth2014apsim}
Dean~P Holzworth, Neil~I Huth, Peter~G deVoil, Eric~J Zurcher, Neville~I Herrmann, Greg McLean, Karine Chenu, Erik~J van Oosterom, Val Snow, Chris Murphy, et~al.
\newblock Apsim--evolution towards a new generation of agricultural systems simulation.
\newblock \emph{Environmental Modelling \& Software}, 62:\penalty0 327--350, 2014.

\bibitem[Basso et~al.(2013)Basso, Cammarano, and Carfagna]{basso2013review}
Bruno Basso, Davide Cammarano, and Elisabetta Carfagna.
\newblock Review of crop yield forecasting methods and early warning systems.
\newblock In \emph{Proceedings of the first meeting of the scientific advisory committee of the global strategy to improve agricultural and rural statistics, FAO Headquarters, Rome, Italy}, volume 241, 2013.

\bibitem[Newlands et~al.(2014)Newlands, Zamar, Kouadio, Zhang, Chipanshi, Potgieter, Toure, and Hill]{newlands2014integrated}
Nathaniel~K Newlands, David~S Zamar, Louis~A Kouadio, Yinsuo Zhang, Aston Chipanshi, Andries Potgieter, Souleymane Toure, and Harvey~SJ Hill.
\newblock An integrated, probabilistic model for improved seasonal forecasting of agricultural crop yield under environmental uncertainty.
\newblock \emph{Frontiers in Environmental Science}, 2:\penalty0 17, 2014.

\bibitem[Bussay et~al.(2015)Bussay, van~der Velde, Fumagalli, and Seguini]{bussay2015improving}
Attila Bussay, Marijn van~der Velde, Davide Fumagalli, and Lorenzo Seguini.
\newblock Improving operational maize yield forecasting in hungary.
\newblock \emph{Agricultural Systems}, 141:\penalty0 94--106, 2015.

\bibitem[Paudel et~al.(2021)Paudel, Boogaard, de~Wit, Janssen, Osinga, Pylianidis, and Athanasiadis]{paudel2021machine}
Dilli Paudel, Hendrik Boogaard, Allard de~Wit, Sander Janssen, Sjoukje Osinga, Christos Pylianidis, and Ioannis~N Athanasiadis.
\newblock Machine learning for large-scale crop yield forecasting.
\newblock \emph{Agricultural Systems}, 187:\penalty0 103016, 2021.

\bibitem[Drummond et~al.(2003)Drummond, Sudduth, Joshi, Birrell, and Kitchen]{site1}
Scott~T Drummond, Kenneth~A Sudduth, Anupam Joshi, Stuart~J Birrell, and Newell~R Kitchen.
\newblock Statistical and neural methods for site--specific yield prediction.
\newblock \emph{Transactions of the ASAE}, 46\penalty0 (1):\penalty0 5, 2003.

\bibitem[Peralta et~al.(2016)Peralta, Assefa, Du, Barden, and Ciampitti]{site2}
Nahuel~R Peralta, Yared Assefa, Juan Du, Charles~J Barden, and Ignacio~A Ciampitti.
\newblock Mid-season high-resolution satellite imagery for forecasting site-specific corn yield.
\newblock \emph{Remote Sensing}, 8\penalty0 (10):\penalty0 848, 2016.

\bibitem[Anselin et~al.(2004)Anselin, Bongiovanni, and Lowenberg-DeBoer]{anselin2004spatial}
Luc Anselin, Rodolfo Bongiovanni, and Jess Lowenberg-DeBoer.
\newblock A spatial econometric approach to the economics of site-specific nitrogen management in corn production.
\newblock \emph{American Journal of Agricultural Economics}, 86\penalty0 (3):\penalty0 675--687, 2004.

\bibitem[Schwalbert et~al.(2018)Schwalbert, Amado, Nieto, Varela, Corassa, Horbe, Rice, Peralta, and Ciampitti]{schwalbert2018forecasting}
Rai~A Schwalbert, Telmo~JC Amado, Luciana Nieto, Sebastian Varela, Geomar~M Corassa, Tiago~AN Horbe, Charles~W Rice, Nahuel~R Peralta, and Ignacio~A Ciampitti.
\newblock Forecasting maize yield at field scale based on high-resolution satellite imagery.
\newblock \emph{Biosystems engineering}, 171:\penalty0 179--192, 2018.

\bibitem[Lambert et~al.(2004)Lambert, Lowenberg-Deboer, and Bongiovanni]{lambert2004comparison}
Dayton~M Lambert, James Lowenberg-Deboer, and Rodolfo Bongiovanni.
\newblock A comparison of four spatial regression models for yield monitor data: A case study from argentina.
\newblock \emph{Precision Agriculture}, 5\penalty0 (6):\penalty0 579--600, 2004.

\bibitem[Li et~al.(2016)Li, Coble, Tack, and Barnett]{li2016estimating}
Xiaofei Li, Keith~H Coble, Jesse~B Tack, and Barry~J Barnett.
\newblock Estimating site-specific crop yield response using varying coefficient models.
\newblock Technical report, 2016.

\bibitem[Shand et~al.(2018)Shand, Li, Park, and Albarrac\'{\i}n]{Shand2018}
Lyndsay Shand, Bo~Li, Trevor Park, and Dolores Albarrac\'{\i}n.
\newblock Spatially varying auto-regressive models for prediction of new human immunodeficiency virus diagnoses.
\newblock \emph{J. R. Stat. Soc. Ser. C. Appl. Stat.}, 67\penalty0 (4):\penalty0 1003--1022, 2018.
\newblock ISSN 0035-9254.
\newblock \doi{10.1111/rssc.12269}.
\newblock URL \url{https://doi.org/10.1111/rssc.12269}.

\bibitem[Perry and Niemann(2007)]{perry2007}
Mark~A Perry and Jeffrey~D Niemann.
\newblock Analysis and estimation of soil moisture at the catchment scale using eofs.
\newblock \emph{Journal of Hydrology}, 334\penalty0 (3-4):\penalty0 388--404, 2007.

\bibitem[Korres et~al.(2010)Korres, Koyama, Fiener, and Schneider]{korres2010}
W~Korres, CN~Koyama, P~Fiener, and K~Schneider.
\newblock Analysis of surface soil moisture patterns in agricultural landscapes using empirical orthogonal functions.
\newblock \emph{Hydrology \& Earth System Sciences}, 14\penalty0 (5), 2010.

\bibitem[Franz et~al.(2020)Franz, Pokal, Gibson, Zhou, Gholizadeh, Tenorio, Rudnick, Heeren, McCabe, Ziliani, et~al.]{franz2020role}
Trenton~E Franz, Sayli Pokal, Justin~P Gibson, Yuzhen Zhou, Hamed Gholizadeh, Fatima~Amor Tenorio, Daran Rudnick, Derek Heeren, Matthew McCabe, Matteo Ziliani, et~al.
\newblock The role of topography, soil, and remotely sensed vegetation condition towards predicting crop yield.
\newblock \emph{Field Crops Research}, 252:\penalty0 107788, 2020.

\bibitem[Eyinla and Oladunjoye(2014)]{ecyield}
DorcasS Eyinla and Michael~A Oladunjoye.
\newblock Improving quality agricultural practices in tropical environments through integrated geophysical methods.
\newblock 2014.

\bibitem[Leroux et~al.(2000)Leroux, Lei, and Breslow]{leroux}
Brian~G. Leroux, Xingye Lei, and Norman Breslow.
\newblock Estimation of disease rates in small areas: a new mixed model for spatial dependence.
\newblock In \emph{Statistical models in epidemiology, the environment, and clinical trials ({M}inneapolis, {MN}, 1997)}, volume 116 of \emph{IMA Vol. Math. Appl.}, pages 179--191. Springer, New York, 2000.
\newblock \doi{10.1007/978-1-4612-1284-3_4}.
\newblock URL \url{https://doi.org/10.1007/978-1-4612-1284-3_4}.

\bibitem[Nielsen(2016)]{Nielsen2016}
Frank Nielsen.
\newblock \emph{Hierarchical Clustering}, pages 195--211.
\newblock Springer International Publishing, Cham, 2016.
\newblock ISBN 978-3-319-21903-5.
\newblock \doi{10.1007/978-3-319-21903-5_8}.
\newblock URL \url{https://doi.org/10.1007/978-3-319-21903-5_8}.

\bibitem[von Luxburg(2007)]{enn}
Ulrike von Luxburg.
\newblock A tutorial on spectral clustering.
\newblock \emph{Stat. Comput.}, 17\penalty0 (4):\penalty0 395--416, 2007.
\newblock ISSN 0960-3174.
\newblock \doi{10.1007/s11222-007-9033-z}.
\newblock URL \url{https://doi.org/10.1007/s11222-007-9033-z}.

\end{thebibliography}

\newpage
\section*{Appendix}
\label{appendix}





\subsection*{A. Missing data}
\label{appendix:Appendix_missingdata}

Consider the normalized yield matrix ${\mathbf{Z}} \sim MVN(\boldsymbol\mu,\boldsymbol\Sigma )$ with $n \times T$ observations. 
Let ${\mathbf{Z}}$ be partitioned as, 
\begin{align*}
    {\mathbf{Z}}={\begin{bmatrix}{\mathbf  {Z}}_{2009}\\{\mathbf  {Z}}_{-2009}\end{bmatrix}}{\text{ with sizes }}{\begin{bmatrix}n\times 1\\n(T-1)\times 1\end{bmatrix}},
\end{align*}
where ${\mathbf{Z_{-2009}}}$ is the matrix of normalized yield for all years except the year 2009. Accordingly, we partition $\boldsymbol\mu$ and $\boldsymbol\Sigma$ as follows,
\begin{align*}
    \boldsymbol\mu
&=
\begin{bmatrix}
 \boldsymbol\mu_1 \\
 \boldsymbol\mu_2
\end{bmatrix}
\text{ with sizes }\begin{bmatrix} n \times 1 \\ n(T-1) \times 1 \end{bmatrix},\\
\boldsymbol\Sigma
&=
\begin{bmatrix}
 \boldsymbol\Sigma_{11} & \boldsymbol\Sigma_{12} \\
 \boldsymbol\Sigma_{21} & \boldsymbol\Sigma_{22}
\end{bmatrix}
\text{ with sizes }\begin{bmatrix} n \times n & n \times n(T-1) \\ n(T-1) \times T & n(T-1) \times n(T-1) \end{bmatrix},
\end{align*}
then the distribution of $\mathbf{Z_{2009}}$ conditional on ${\mathbf{Z_{-2009}}} = 
\mathbf{a}$ is multivariate normal given by, $( {\mathbf{Z_{2009}}} | {\mathbf{Z_{-2009}}}= \mathbf a) \sim N(\bar{\boldsymbol\mu}, \bar{\boldsymbol\Sigma})$, where
\begin{align*}
  {\bar{\boldsymbol{\mu}}} = \boldsymbol\mu_1 + 
\boldsymbol\Sigma_{12} \boldsymbol\Sigma_{22}^{-1}
\left( \mathbf{a} - \boldsymbol\mu_2 \right), \ \text{and} \ {\overline {\boldsymbol {\Sigma }}}={\boldsymbol {\Sigma }}_{11}-{\boldsymbol {\Sigma }}_{12}{\boldsymbol {\Sigma }}_{22}^{-1}{\boldsymbol {\Sigma }}_{21}.  
\end{align*}
We update the estimates of $\bar{\boldsymbol\mu}$ and $\bar{\boldsymbol\Sigma}$ as one of the steps in the MCMC algorithm using Gibbs sampling. We then take a random sample from N($\bar{\boldsymbol\mu}$, $\bar{\boldsymbol\Sigma}$) and update the $\mathbf{Z}$ matrix in each step of the MCMC algorithm. 
\\

\subsection*{B. Results for different values of $\epsilon$}
\label{appendix:Appendix_epsresults_mead}

We present the forecasting results obtained for different values of $\epsilon$ in Table \ref{tab:epsilon}.
\begin{itemize}
    \item Case 1: Some of the clusters have no neighbors ($\epsilon = 3$),
    \item Case 2: Every cluster has at least one neighbor ($\epsilon =  6$),
    \item Case 3: Every cluster has at least two neighbors ($\epsilon =  16$).
\end{itemize}
At the Mead site, we have observed that the outcomes for Cases when $\epsilon = $ 6 and 16 are analogous to those for $\epsilon = 100$, where all clusters are considered neighbors of each other. However, the forecasting performance is slightly reduced for Case when $\epsilon = 3$, since certain clusters have no neighbors. Nevertheless, overall, the results are not sensitive to the value of $\epsilon$, given that each cluster has at least one neighbor.

For the blocking-based SVAR model, we have also considered the exchangeable prior proposed in \cite{Shand2018} to determine if it can enhance the forecasting accuracy of the model. The exchangeable prior assumes that all blocks are neighbors of each other. The outcomes are shown in Table \ref{tab:epsilon}. However, using the exchangeable prior for the neighborhood matrix structure produces only minor improvements in the forecasting performance compared to using the spatial neighborhood matrix structure which considers the blocks on the border as neighbors.

\begin{table}[H] 
\caption{\label{tab:epsilon} Forecasting performance at the Mead site for different values of $\epsilon$} 
\centering 
\scalebox{0.9}{
\renewcommand{\arraystretch}{2}
\begin{tabular}{p{1.7cm} p{1.6cm} c c  c  c  c  p{1.7cm} } 
\hline\hline  
\textbf{Aggregation Method} & \textbf{Forecasting Model} & \textbf{\# Clusters} & \textbf{$\epsilon$}  & \textbf{$R^2\,(\%)$} & \textbf{MSPE} & \textbf{MAPE} & \textbf{Predicted Average} \\ [0.5ex] 

\hline 
Clustering & SVAR          & 25 &100 & 81      & 0.141 & 0.322 & 11.717 \\
Clustering & SVAR          & 25 &3 & 79.609	    & 0.151 & 0.327 & 11.69 \\
Clustering & SVAR          & 25 &6 & 80.741      & 0.143 & 0.323 & 11.706 \\
Clustering & SVAR          & 25 &16 & 80.924      & 0.142 & 0.324 & 11.713 \\

\hline
Blocking   & SVAR          & 25 & Spatial NB & 28.637  & 0.316 & 0.461 & 11.693 \\
Blocking   & SVAR          & 25 & Exchangeable & 31.244  & 0.304 & 0.456 & 11.705 \\

\hline 
\end{tabular}
}
\end{table}

\subsection*{C. Results for Brule and Site 6}
\label{appendix:Appendix_other_site_results}

\begin{figure}[H]
  \includegraphics[scale=0.34]{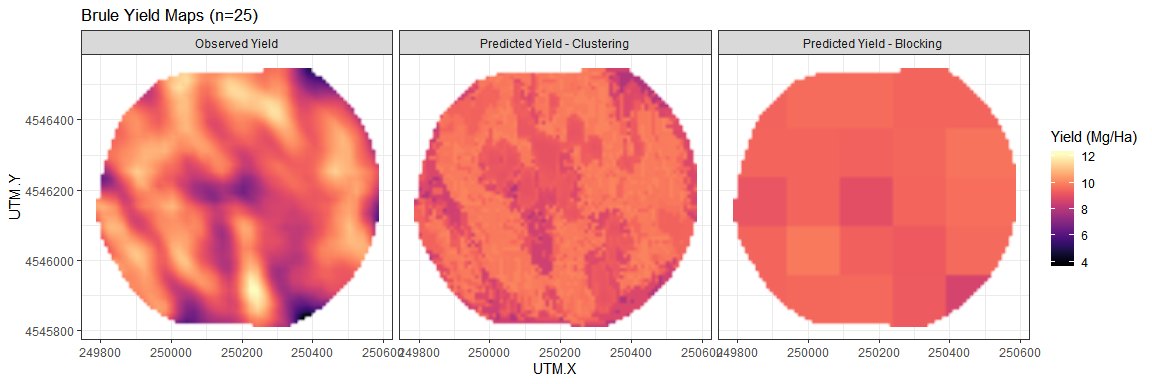}
  \caption{Yield maps for Site Brule for 25 clusters, A: Observed yield for 2017, B: Predicted yield for 2017 using clustering based SVAR model, C: Predicted yield for 2017 using blocking based SVAR model}
  \label{fig:Brule_yieldmaps_25}
\end{figure}

\begin{figure}[H]
  \includegraphics[scale=0.34]{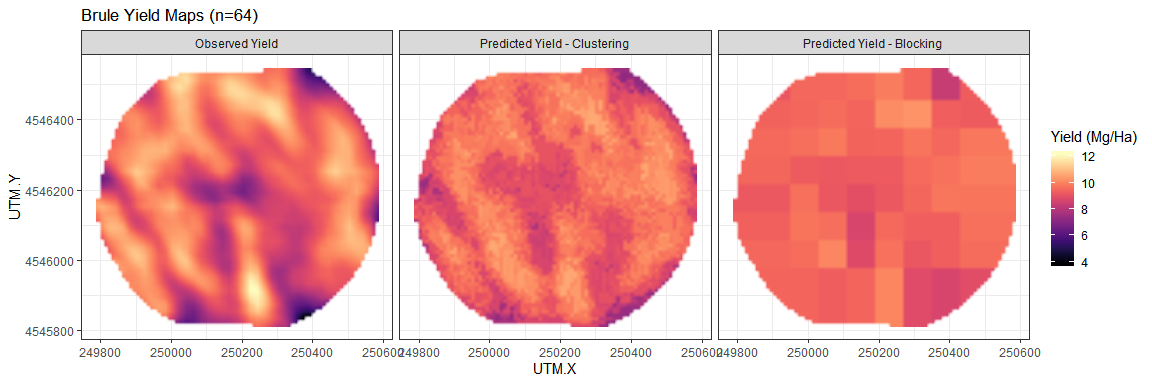}
  \caption{Yield maps for Site Brule for 64 clusters, A: Observed yield for 2017, B: Predicted yield for 2017 using clustering based SVAR model, C: Predicted yield for 2017 using blocking based SVAR model}
  \label{fig:Brule_yieldmaps_64}
\end{figure}

\begin{figure}[H]
  \includegraphics[scale=0.34]{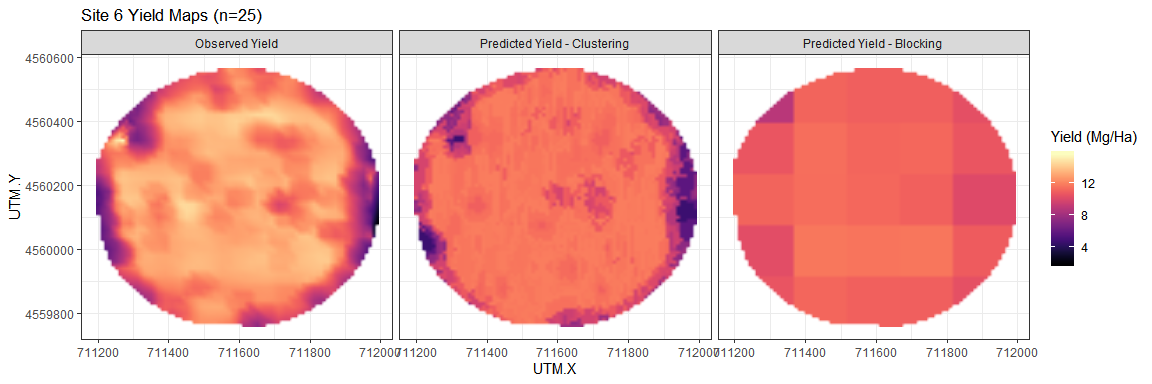}
  \caption{Yield maps for Site 6 for 25 clusters, A: Observed yield for 2017, B: Predicted yield for 2017 using clustering based SVAR model, C: Predicted yield for 2017 using blocking based SVAR model}
  \label{fig:Site6_yieldmaps_25}
\end{figure}

\begin{figure}[H]
  \includegraphics[scale=0.34]{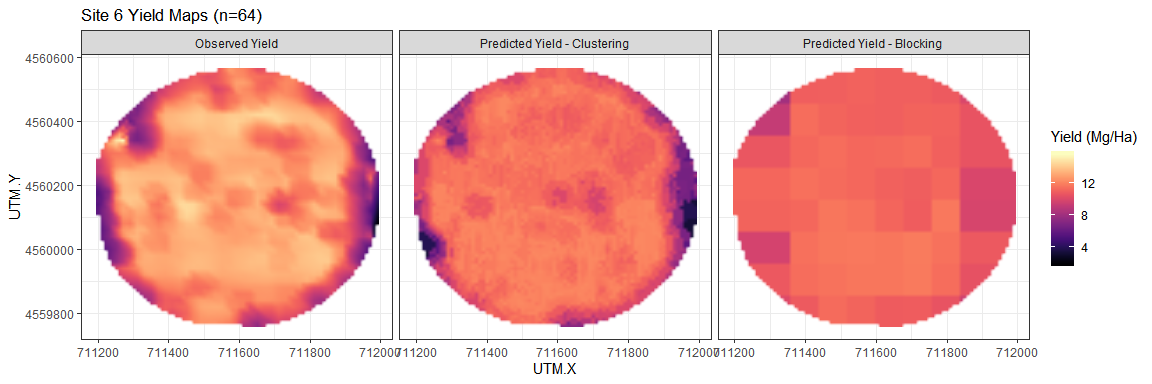}
  \caption{Yield maps for Site 6 for 64 clusters, A: Observed yield for 2017, B: Predicted yield for 2017 using clustering based SVAR model, C: Predicted yield for 2017 using blocking based SVAR model}
  \label{fig:Site6_yieldmaps_64}
\end{figure}

We present the forecasting performance for two independent sites, Brule and Site 6, along with the yield maps generated using $25$ and $64$ clusters. The results, which are illustrated in Table \ref{Forecasting_performance_table_Brule} and \ref{Forecasting_performance_table_Site_6}, and Figure \ref{fig:Brule_yieldmaps_25}, \ref{fig:Brule_yieldmaps_64}, \ref{fig:Site6_yieldmaps_25}, \ref{fig:Site6_yieldmaps_64}, demonstrate that the clustering-based SVAR model with $64$ clusters outperforms other models in terms of forecasting accuracy and quality of yield maps.

\begin{table}[H]
\caption{Forecasting performance at the Site Brule} 
\label{Forecasting_performance_table_Brule}
\centering 
\scalebox{0.9}{
\renewcommand{\arraystretch}{2}
\begin{tabular}{p{1.7cm} p{1.6cm} c  c  c  c  p{1.7cm} } 
\hline\hline  
\textbf{Aggregation Method} & \textbf{Forecasting Model} & \textbf{\# Clusters}  & \textbf{$R^2\,(\%)$} & \textbf{MSPE} & \textbf{MAPE} & \textbf{Predicted Average} \\ [0.5ex] 
\hline 
Clustering & SVAR          & \multirow{2}{*}{25} & 85.252  & 0.085 & 0.255 & 9.48 \\
Clustering & Random Forest &                     & 93.227  & 0.039 & 0.154 & 9.51 \\
\hline
Blocking   & SVAR          & \multirow{2}{*}{25} & 35.892  & 0.273 & 0.447 & 9.482 \\
Blocking   & Random Forest &                     & 66.416  & 0.143 & 0.323 & 9.507 \\
\hline
Clustering & SVAR          & \multirow{2}{*}{64} & 90.447  & 0.1   & 0.251 & 9.497 \\
Clustering & Random Forest &                     & 91.256  & 0.092 & 0.202 & 9.508 \\
\hline
Blocking   & SVAR          & \multirow{2}{*}{64} & 47.652  & 0.4   & 0.511 & 9.496 \\
Blocking   & Random Forest &                     & 74.309  & 0.196 & 0.348 & 9.497 \\
\hline
None       & Random Forest &         -           & 73.577  & 0.32  & 0.437 & 9.487 \\ 
\hline 
\end{tabular}
}
\end{table}
\newpage

\begin{table}[H]
\caption{Forecasting performance at the Site 6} 
\label{Forecasting_performance_table_Site_6}
\centering 
\scalebox{0.9}{
\renewcommand{\arraystretch}{2}
\begin{tabular}{p{1.7cm} p{1.6cm} c  c  c  c  p{1.7cm} } 
\hline\hline  
\textbf{Aggregation Method} & \textbf{Forecasting Model} & \textbf{\# Clusters}  & \textbf{$R^2\,(\%)$} & \textbf{MSPE} & \textbf{MAPE} & \textbf{Predicted Average} \\ [0.5ex] 
\hline 
Clustering & SVAR          & \multirow{2}{*}{25} & 74.188  & 1.044 & 0.915 & 11.047 \\
Clustering & Random Forest &                     & 69.083  & 1.25  & 1.037 & 11.079 \\
\hline
Blocking   & SVAR          & \multirow{2}{*}{25} & 24.099  & 1.238 & 0.962 & 11.064 \\
Blocking   & Random Forest &                     & 10.628  & 1.457 & 1.051 & 11.053 \\
\hline
Clustering & SVAR          & \multirow{2}{*}{64} & 75.641  & 1.105 & 0.943 & 11.044 \\
Clustering & Random Forest &                     & 71.386  & 1.298 & 1.037 & 11.073 \\
\hline
Blocking   & SVAR          & \multirow{2}{*}{64} & 21.192  & 1.623 & 1.078 & 11.076 \\
Blocking   & Random Forest &                     & 8.555   & 1.883 & 1.183 & 11.039 \\
\hline
None       & Random Forest &         -           & -97.876  & 6.922 & 2.403 & 9.69 \\ 
\hline 
\end{tabular}
}
\end{table}
\newpage

\end{document}